\documentclass[twocolumn,pra,aps,showpacs,superscriptaddress]{revtex4}
\usepackage{amsmath,amssymb,amsfonts,bbm,graphicx}
\usepackage{amstext}
\usepackage{color}
\usepackage{soul}

\renewcommand{\[}{\begin{equation}}
\renewcommand{\]}{\end{equation}}

\begin{document}

\title{Tunneling, self-trapping and manipulation of higher modes of a BEC in a double well}

\author{J.~Gillet}
\email{jeremie.gillet@oist.jp}
\affiliation{Department of Physics, University College Cork, Cork, Ireland}
\affiliation{Okinawa Institute of Science and Technology Graduate University, Onna-son, Okinawa, Japan}
\author{M.~A.~Garcia-March}
\affiliation{Department of Physics, University College Cork, Cork, Ireland}
\affiliation{Department d'Estructura i constituents de la Materia, Universitat de Barcelona, Barcelona, Spain}
\author{Th.~Busch}
\affiliation{Department of Physics, University College Cork, Cork, Ireland}
\affiliation{Okinawa Institute of Science and Technology Graduate University, Onna-son, Okinawa, Japan}
\author{F.~Sols}
\affiliation{Departamento de F\'isica de Materiales, Universidad Complutense de Madrid, E-28040 Madrid, Spain}

\date{\today}

\begin{abstract}
We consider an atomic Bose-Einstein condensate trapped in a symmetric one-dimensional double well potential in the four-mode approximation and show that the semiclassical dynamics of the two ground state modes  can be strongly influenced by a macroscopic occupation of the two excited modes. In particular, the addition of the two excited  modes already unveils features related to the effect of dissipation on the condensate.  In general, we find a rich dynamics that includes Rabi oscillations, a mixed Josephson-Rabi regime, self-trapping, chaotic behavior, and the existence of fixed points.
We  investigate how the dynamics of the atoms in the excited modes can be manipulated  by controlling the atomic populations of the ground states.
\end{abstract}


\maketitle

\section{Introduction}

Recent experimental realizations of systems of ultracold atoms prepared in excited Bloch bands in optical lattices have opened new prospects in ultracold atomic science~\cite{Browaeys:2005,Spielman:2006,Muller:2007,Clement:2009,Wirth:2011,Olschlager:2011}. This access to the orbital degree of freedom allows to observe new exotic physics by playing with the anisotropy of the Wannier functions from which the Bloch states of the excited bands are built. In particular, they can possess new quantum degeneracies associated with the symmetries of the system~\cite{Lewenstein:2011}. The population of the excited levels was also shown to be important in the study of the simplest building block of an optical lattice, the double well. Here, the excited levels are responsible for enhancing the tunneling of atoms through the barrier~\cite{Chatterjee:2010,JuliaDiaz:2010b}, which is a process that has been suggested for the creation of macroscopic superposition states with orbital degrees of
freedom and two-qubit phase gates~\cite {Strauch:2008,GarciaMarch:2011,GarciaMarch:2012}. Also, the Josephson effect between different orbital states within the same region of space was predicted to exist in externally driven condensates~\cite{Heimsoth:2012}.

\begin{figure}
\includegraphics[width=\linewidth]{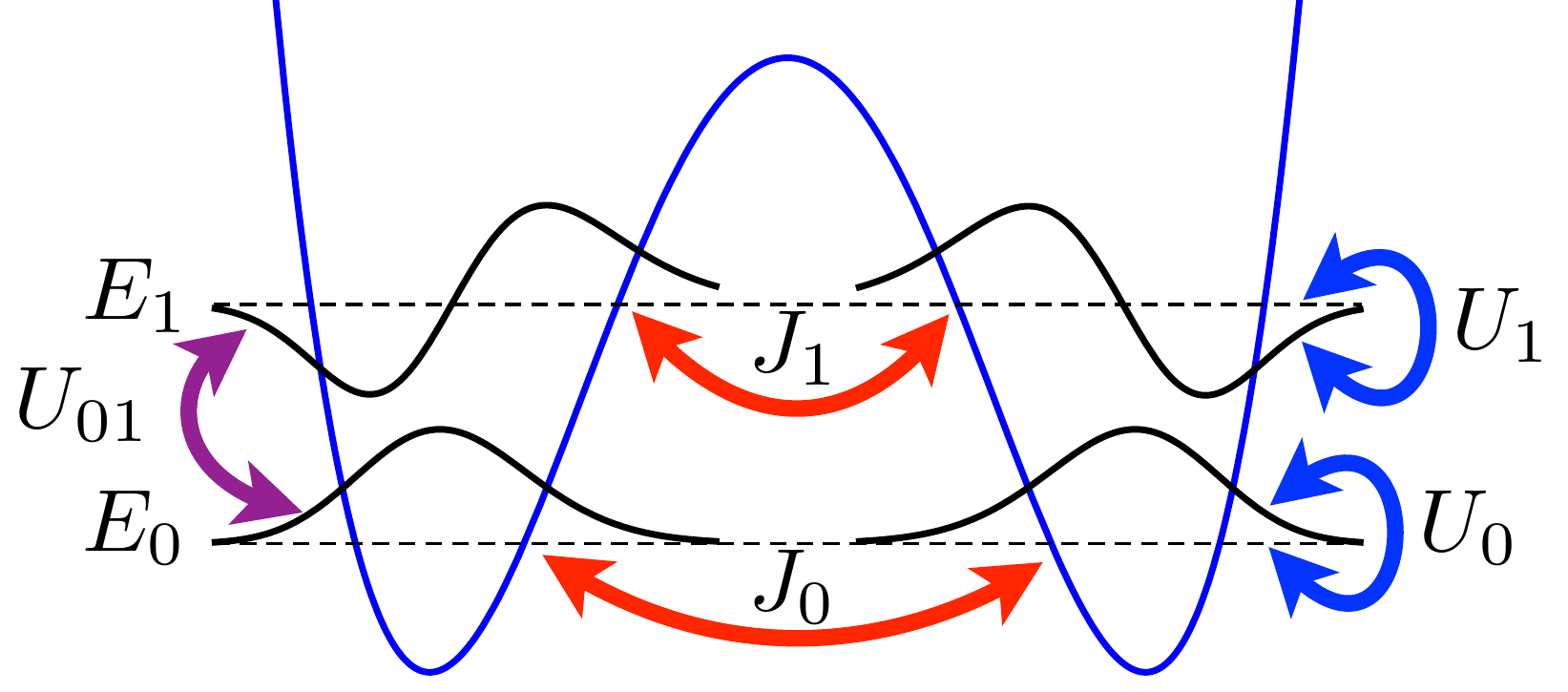}
   \caption{(Color online)  {\it Schematics of the symmetric double well.} The first four modes of energy $E_0$ and $E_1$ are represented. The $J_\ell$ terms ($\ell=0,1$) indicate the  tunneling energy in each level. The $U_\ell$ terms represent the interaction energy in each level. The $U_{01}$ term is the inter-level interaction energy.}
\label{fig:DW}
\end{figure}

Ultracold atoms in double wells have been thoroughly studied in the framework of two-mode models, both with an eye towards unveiling Josephson physics~\cite{Javanainen:1987,Smerzi:1997,Milburn:1997,Zapata:1998,Raghavan:1999,Ostrovskaya:2000,Mahmud:2005,Ananikian:2006,Fu:2006,JuliaDiaz:2010,Albiez:2005,Levy:2007,Zibold:2010, Dalton:2012,Gertjerenken:2013b,Jezek:2013,Mazzarella:2013} and as a candidate system to observe macroscopic atomic quantum superpositions~\cite{Steel:1998,Cirac:1998,Gordon:1999,Higbie:2004,Huang:2006, Piazza:2008,Mazets:2008,Carr:2010,Watanabe:2010,He:2012,Csire:2012,Gertjerenken:2013}. In this manuscript, we introduce a semiclassical approach that allows one to study effects on the double-well condensate dynamics stemming from a finite population of the excited single-particle states (see Fig.~\ref{fig:DW}). We study the situation where the presence of excited states yields a simplified model for the effect of dissipation on the collective dynamics of atoms oscillating
coherently between the ground states of the two wells.
We also consider the physically realizable case where a significant part of the gas is intentionally excited to the higher lying states.

Two-mode approaches are conventionally used to model ultracold bosons in double wells \cite{Javanainen:1987,Smerzi:1997,Milburn:1997,Zapata:1998,Raghavan:1999,Ostrovskaya:2000,Mahmud:2005,Ananikian:2006,Fu:2006,JuliaDiaz:2010, Dalton:2012,Gertjerenken:2013b,Jezek:2013,Mazzarella:2013}. The relevant parameters in such a model are the tunneling energy $J$ and the interaction energy $U$, together with the total number of atoms $N$. According
to which energy  dominates one can identify three main regimes \cite{Sols:1999b,Legget:2001}. For the  Rabi regime ($U \ll J/N$), macroscopic tunneling of essentially independent atoms between both wells is predicted. Increasing the atom-atom interaction ($J/N \ll U \ll NJ$), the system will enter the  Josephson regime in which, for certain initial conditions, macroscopic self-trapping is possible~\cite{Smerzi:1997,Milburn:1997,Zapata:1998,Raghavan:1999}. Finally, the Fock regime is reached ($NJ \ll U$),  in which semiclassical approaches cease to be adequate.

Several methods exists that allow to theoretically treat the double well: the semiclassical approximation maps its dynamics onto a
non-rigid pendulum~\cite{Smerzi:1997,Milburn:1997,Raghavan:1999} and in the strong correlation regime direct diagonalization of the  two-site Bose-Hubbard Hamiltonian (originally called the  Lipkin-Meshkov-Glick Hamiltonian in nuclear physics~\cite{Lipkin:1965,Vidal:2004}) can be undertaken. The correspondence between the exact two-mode many-boson Hamiltonian and the semiclassical approach can be elucidated by means of a phase-space distribution function. In this context, the  Husimi distribution function has been used to express the quantum results in a semiclassical language~\cite{Mahmud:2005}. An interesting  outcome is that, while in the Josephson regime the quantum approach always maintains the parity of the Hamiltonian, the semiclassical approach breaks that symmetry, allowing in particular for the phenomenon of macroscopic self-trapping~\cite{Mahmud:2005}. One of the numerical methods that have successfully been used to study this system within the quantum approach is, for
example, the
multiconfigurational time-dependent Hartree methods (MCTDH)~\cite{Masiello:2005,Streltsov:2006,Alon:2008,Sakmann:2009}.

Here we extend the two-mode model to include two additional excited modes. This extension is motivated by several reasons. It is  a first step towards a complete interpolation between two limits in the dynamics of $N$ atoms (with $N$ large): (i) the dimension of the one-atom Hilbert space is two (which amounts to an $N$-body extension of  the two-level system), and (ii) the dimension is $M\sim N$ (which resembles the problem of ultracold atoms in optical lattices). On the other hand, it will allow us to test the validity and limits of the common two-mode approach in more detail. It will also help in benchmarking the more exact but numerically demanding MCTDH approach. Another important motivation for this work is the goal of mimicking the effect of dissipation through the existence of  higher lying states into which the atoms can be excited, in what could be viewed as a toy model of the depletion cloud. We will argue that some aspects of real dissipation can already be explored within this simplified 
description.

Finally, a further motivation is  to show that the presence of higher modes can lead to new double-well dynamics which is interesting in its own right. Particularly, we will see that the short time dynamics  of the excited modes can be manipulated by means of  the initial population imbalance of the ground modes. This is quite relevant for the current state of the art, as recent experimental progress has shown that it is possible to invert the populations in a single well (see Refs.~\onlinecite{Bucker:2011,Bucker:2012}).

The semiclassical limit is described by  two non-rigid pendula for which the canonically conjugate variables are  the population imbalances  and phase differences between left and right wells for both the ground and excited modes.  These pendula are coupled by a third pendulum,  for which the canonically conjugate variables are number and phase difference between the total populations (i.e., summed over the two wells)  in the excited and ground modes.
This latter pendulum stems from the local interaction that transfers atoms between the ground and the excited modes. In the Rabi regime, the dynamics of the first two pendula is similar to that of two uncoupled pendula. By contrast, the population imbalance associated to the third pendulum remains constant, while the corresponding phase grows unbounded. This aspect of the non-interacting picture remains qualitatively valid in the Josephson regime for a considerable time during which the population imbalance between  ground and excited modes remains approximately constant. We note, however, that actual fixed points for this third pendulum exist only in the Fock regime, which lies outside the limits of applicability of the semiclassical model.
Indeed, the nominally opposite extreme of negligible interactions  is the limit of an infinitely high barrier.  In that limit, the population imbalances in the first two pendula are of course self-trapped, but the local interaction can still be viewed as causing  intra-well tunneling of atoms between the ground and excited modes.

From our semiclassical study we find that, in addition to the known bifurcation at the fixed point in the transition from the Rabi to the Josephson regime~\cite{Zibold:2010},  new fixed points emerge which stem from the competing effects of  inter-well tunneling and  intra-well inter-level (local) interactions. We find that the inclusion of additional modes leads to  a shift in the tunneling frequencies. We also show that in the Josephson regime the population of the ground modes can be used  to control the tunneling dynamics of the excited ones.

The paper is organized as follows. In Sec.~\ref{sec2} we introduce the many-body Hamiltonian and in Sec.~\ref{sec3} we formulate the semiclassical model. In Sec.~\ref{sec4} we describe the various regimes, including the limits of validity of the model and the stability of the fixed points. Section \ref{sec6} deals with the dynamics of the four-mode problem,  with an emphasis on the frequencies shifts and their relation to dissipation. In Sec.~\ref{sec7} we focus on the control of the excited-mode tunneling dynamics through the  population imbalance of the ground modes. Conclusions are presented in Sec.~\ref{sec8}.

\section{Theoretical Model}
\label{sec2}

Let us consider $N$ interacting bosons of mass $M$ confined by an external potential $V(\mathbf{r})$.  We assume point-like contact interactions, corresponding to low densities and energies, with a coupling strength $g$ depending on the $s$-wave scattering length $a_s$ of the atoms, $g=4\pi \hbar^2 a_s/M$. Let us consider the anisotropic case in which we squeeze the trap in the $x$ and $y$ directions, while the potential in the $z$ direction resembles that of a double well
\begin{equation}
    V(\mathbf{r})=\frac{1}{2}M\,w_{\perp}^2(x^2+y^2)+ V_{\mathrm{dw}}(z).
\end{equation}
We assume that the oscillator length in the transverse direction, $a_{\mathrm{ho},\perp}\equiv\sqrt{\hbar/M\omega_{\perp}}$ satisfies
$a_{\mathrm{ho},\perp}\ll a_s$, so the system is in the tight confinement limit, and can be considered effectively one dimensional~\cite{Olshanii:1998}. Then, the contact interactions are governed by the coupling constant  $g_{1}=-2\pi\hbar^{2}/M\,a_{1D}$, with
$a_{1D}=(-(a_{\mathrm{ho},\perp})^{2}/2a_s)[1-C(a_s/a_{\mathrm{ho},\perp})]$ and $C\simeq 1.4603$~\cite{Olshanii:1998}. Without loss of generality, we consider a particular type of double-well potential given by the Duffing form
\begin{equation}
\label{eq:Duffing}
    V_{\mathrm{dw}}(z)=V_{0}\left[1-4\left(\frac{z}{a}\right)^2\right]^2.
\end{equation}
It is convenient at this point to renormalize the spatial coordinate to be dimensionless, $\tilde z\equiv z/a$, and hereafter we use the notation $z$ when actually referring to $\tilde z$.

We consider four spatial modes represented by the functions $\psi_{j\ell}(z)$, for which the index $j\in\{L,R\}$ accounts for the atoms localized in either well while the index $\ell\in\{0,1\}$ indicates occupation of the ground or the first excited energy
level. The Hamiltonian can then be written as
\begin{equation}
   H=H_{0}+H_{1}+H_{01},
   \label{eq:two-level}
\end{equation}
where
\begin{align}
  H_{\ell} =&E_{\ell}\left(n_{L\ell}+n_{R\ell}\right)
                     -J_{\ell} \left(b_{L\ell}^{\dagger}b_{R\ell}+b_{R\ell}^{\dagger}b_{L\ell}\right)\nonumber \\
                  &\,+U_{\ell}\sum_{j}n_{j\ell}\left(n_{j\ell}-1\right),\label{eq:two-levela}
\end{align}
and
\begin{equation}
   H_{01}=U_{01}\sum_{j,\ell\ne \ell'}\left(2n_{j\ell}n_{j\ell'}+b_{j\ell}^{\dagger}b_{j\ell}^{\dagger}b_{j\ell'}b_{j\ell'}\right).
   \label{eq:two-levelb}
\end{equation}
In Eqs.~(\ref{eq:two-level})-(\ref{eq:two-levelb}), $n_{j\ell}=b_{j\ell}^{\dagger}b_{j\ell}$, and $b_{j\ell}$ and $b_{j\ell}^{\dagger}$ are the annihilation and creation operators satisfying the usual bosonic commutation relations. The level, hopping, and interaction energies are given by
\begin{align}
   E_{\ell}\equiv&\int\psi_{j\ell}^{\ast}(z)H_{\mathrm{sp}}\psi_{j\ell}(z)\, dz,\label{eq:E_lm} \\
   J_{\ell}\equiv&- \int\psi_{j\ell}^{\ast}(z)H_{\mathrm{sp}}\psi_{j'\ell}(z)\, dz,\label{eq:J_lm} \\
   U_{\ell}\equiv&\frac{g_{1}}{2}\int|\psi_{j\ell}(z)|^{4}\, dz,\label{eq:U_l}\\
   U_{01}\equiv&\frac{g_{1}}{2}\int|\psi_{j0}(z)|^{2}|\psi_{j1}(z)|^{2}\, dz,\label{eq:U_lm_l'm'}
\end{align}
with
\begin{equation}
\label{Eq:spH}
   H_{\mathrm{sp}}\equiv-\frac{\hbar^{2}}{2M}\partial_{zz}+V_{\mathrm{dw}}(z),
\end{equation}
the single-particle Hamiltonian. Note that in Eq.~(\ref{eq:J_lm}) the value of the coefficients does not depend on the choice of $j\neq j'$ due to the symmetry of the potential. The Hamiltonian~(\ref{eq:two-level})-(\ref{eq:two-levelb}) conserves the total number of atoms, $\sum_{j\ell}N_{j\ell}$, and neglects processes whereby an atom changes well and level simultaneously. The characteristic energies associated to such processes are typically at least two orders of magnitude  smaller than $U_{01}$~\cite{GarciaMarch:2012b,Spekkens:1998}. A diagram of the double well with all the interaction terms and hopping terms considered is shown in Fig.~\ref{fig:DW}.

The recoil energy associated with a 1D periodic optical lattice of wavelength $\lambda$ is defined as $E_r = 2\hbar^2 \pi^2 / M\lambda^2$. Since  $\lambda \sim a$, we substitute $\lambda$ by $a$ and, dividing the potential by $E_r$, we can write $V_{\mathrm{dw}}(z)/E_r=\tilde{V}_0(1-4z^2)^2$ with the dimensionless strength
\begin{equation}
   \tilde{V}_0\equiv\frac{Ma^2V_0}{2\hbar^2 \pi^2 }.
\end{equation}
Similarly, we write
\begin{equation}
  \frac{U_{\ell}}{E_r}=\frac{\tilde{g}_{1}}{2}\int|\psi_{j\ell}(z)|^{4}\, dz,
\end{equation}
with
\begin{equation}
   \tilde{g}_1\equiv\frac{g_1Ma}{4\hbar^2 \pi^2 }.
\end{equation}
For readability, in the rest of the paper we will use $V_0$ and $g_1$ when actually referring to $\tilde{V}_0$ and $\tilde{g}_1$. We can then use the two (now dimensionless) parameters $V_0$ and $g_1$ along with the number of atoms $N$ to fully characterize the problem and to numerically compute the functions $\psi_{j\ell}(z)$. The distance $a$ between the wells is included in the scaling
procedure.

\section{Semiclassical approximation}
\label{sec3}

In the few-atom limit and for small interactions, the functions $\psi_{j\ell}(z)$ do not depart significantly from the non-interacting single-particle eigenstates of the individual wells. The spectrum of Hamiltonian~(\ref{eq:two-level})  was thoroughly studied in this few-atom limit in Refs.~\onlinecite{Dounas-Frazer:2007,GarciaMarch:2011,GarciaMarch:2012b}. It was noted there that, once the interaction energies in a single well become larger than the difference between the ground and the first excited state, more than two levels need to be considered. While in this range the exact functional form of the eigenstates changes, the physics discussed below does not depend on these details. Therefore we can simplify the numerics by using the non-interacting eigenfunctions of the Duffing potential to approximate $\psi_{j\ell}(z)$ and to calculate the coefficients~(\ref{eq:E_lm})-(\ref{eq:U_lm_l'm'}). The four-mode approach permits a simplified study of systems in which the two ground state modes coexist with a 
significant
depletion cloud or, more realistically, where a large part of the gas is intentionally excited to the first excited level~\cite{Browaeys:2005,Spielman:2006,Muller:2007,Clement:2009,Wirth:2011,Olschlager:2011,Bucker:2011,Bucker:2012}.

To derive the semiclassical model we start by considering the equations of motion of the destruction operators
\begin{equation}
  - i\frac{db_{j\ell}}{dt}=\left[b_{j\ell},H\right],
\end{equation}
which explicitly  read
\begin{align}
  - i\dot{b}_{j\ell} =& \left( E_{\ell}+ 2 U_{\ell} n_{j\ell} +4U_{01} n_{j\ell'} \right)b_{j\ell}-J_{\ell}b_{j'\ell}\nonumber\\
                            & +2U_{01}b_{j\ell}^{\dagger}b_{j\ell'}^2,
\label{eq:dotb}
\end{align}
where $j'$ stands for $R(L)$ if $j=L(R)$, and similarly for $\ell'$.

Hereafter we focus on the case where the atom number in each mode is large. Then, we can consider $b_{j\ell}$ as the amplitude associated with the mode of wave function $\psi_{j\ell}(z)$. We may write~\cite{davismj2008},
\begin{equation}
  \label{eq:cnum}
   b_{j\ell}=\sqrt{N_{j\ell}}e^{i\phi_{j\ell}}.
\end{equation}
 Standard manipulations lead to
\begin{align}
  \dot{N}_{j\ell}=&-2J_{\ell}\sqrt{N_{j\ell}N_{j'\ell}}\sin\left(\phi_{j\ell}-\phi_{j'\ell}\right) \nonumber\\
                         &+4U_{01}N_{j\ell}N_{j\ell'}\sin\left[2\left(\phi_{j\ell}-\phi_{j\ell'}\right)\right],
\end{align}
\begin{align}
  \dot{\phi}_{j\ell}=&\left(2U_{\ell}N_{j\ell}+E_{\ell}+4U_{01}N_{j\ell'}\right)\nonumber\\
                             &-J_{\ell}\sqrt{\frac{N_{j'\ell}}{N_{j\ell}}}\cos\left(\phi_{j\ell}-\phi_{j'\ell}\right)\nonumber\\
                             &+2U_{01}N_{j\ell'}\cos\left[2\left(\phi_{j\ell}-\phi_{j\ell'}\right)\right].
\end{align}
These equations can be regarded as the equations of motion associated with a classical Hamiltonian $H$ in terms of the canonically conjugated variables $\phi_{j\ell}$ and $N_{j\ell}$, so that $\dot{N}_{j\ell}=-\partial H/\partial\phi_{j\ell}$ and $\dot{\phi}_{j\ell}=\partial H/\partial N_{j\ell}$. We obtain
\begin{align}
      H= \sum_{j\ell}\Big[& \left( E_{\ell} + U_\ell N_{j\ell}\right)N_{j\ell} \nonumber \\
          & -J_{\ell}\sqrt{N_{j\ell}N_{j'\ell}}\cos\left(\phi_{j\ell}-\phi_{j'\ell}\right)  \nonumber \\
          & +U_{01}\left\{2+\cos\left[2\left(\phi_{j\ell}-\phi_{j\ell'}\right)\right]\right\} N_{j\ell}N_{j\ell'} \Big].\label{eq:H0}
\end{align}
As expected, this Hamiltonian conserves the total number of atoms $N=\sum_{j\ell}N_{j\ell}$, since it is independent of the total phase $\theta_N=\sum_{j\ell} \phi_{j\ell}$. We can take advantage of this conservation law and introduce a transformation that reduces the number of dynamical variables to 6 instead of 8:
\begin{equation}
  \begin{bmatrix} 1 \\  z_0(t) \\  z_1(t) \\  z_2(t)\end{bmatrix} =
   \frac{\mathsf{M}}{N} \begin{bmatrix} N_{L,0}(t) \\ N_{R,0}(t) \\ N_{L,1}(t) \\ N_{R,1}(t)\end{bmatrix},
\end{equation}
\begin{equation}
  \label{defoftheta}
  \begin{bmatrix} \theta_N(t) \\ \theta_0(t) \\ \theta_1(t) \\ \theta_2(t)\end{bmatrix} =
  -\mathsf{M}\begin{bmatrix} \phi_{L,0} (t)\\ \phi_{R,0}(t) \\ \phi_{L,1}(t) \\ \phi_{R,1}(t) \end{bmatrix},
\end{equation}
where $\mathsf{M}$ is a $4\times 4$ matrix made of real row vectors orthogonal to each other, with the first of them entirely composed of 1's so that the first variable of the set is the constant number of atoms, $N$. The specific form of $\mathsf{M}$ which defines the three variables $\{z_i,\theta_i\}$ ($i=0,1,2$) can be chosen according to convenience.

This set of new variables $\{z_i,\theta_i\}$ will be canonically conjugate if its Poisson brackets fulfill
\begin{equation}
  \{z_i,\theta_k\} \equiv \sum_{j,\ell} \left( \frac{\partial z_i}{\partial N_{j\ell}} \frac{\partial \theta_k}{\partial \phi_{j\ell}} - \frac{\partial    z_i}{\partial \phi_{j\ell}} \frac{\partial \theta_k}{\partial N_{j\ell}} \right)=\delta_{ik}.
\end{equation}

Let us consider the particular basis given by the transformation matrix
\begin{equation}
   \mathsf{M}=\begin{bmatrix} 1&1&1&1 \\ 1&-1&0&0 \\ 0&0&1&-1 \\ 1&1&-1&-1\end{bmatrix}.
\end{equation}
Note that we have chosen a minus sign to transform the angular variables~\cite{Legget:2001}.  This transformation allows us to express the Hamiltonian in terms of the populations and phase differences between the left and right well for each level
\begin{equation}
 \label{eq:defz}
    z_{\ell}=(N_{L,\ell}-N_{R,\ell})/N,\quad\theta_{\ell}=\phi_{R,\ell}-\phi_{L,\ell}.
\end{equation}
In addition to these four variables, we also use  $z_2$, which is the difference of the total population of the ground and excited modes
\begin{align}
    z_2=&[(N_{L,0}+N_{R,0})-(N_{L,1}+N_{R,1})]/N,\nonumber\\
   \theta_2=&(\phi_{L,1}+\phi_{R,1})-(\phi_{L,0}+\phi_{R,0}).
\end{align}
Due to the constraints on the populations ($0<N_{j\ell}<N$), the range of values that each $z_\ell$ can take is limited  to
\begin{equation}
\label{eq:boundsforzell}
   |z_\ell|<[1+(-1)^\ell z_2]/2
\end{equation}
with $-1<z_2<1$. For example,  if half the atoms in the double well are excited, we have $z_2=0$ and both $z_\ell$ vary between $-1/2$ and $1/2$. 

In this new basis, the renormalized classical Hamiltonian $H'=2H/N-E_0-E_1$, follows from Eq.~(\ref{eq:H0}) and takes the form
\begin{eqnarray}
H' &=& - J_0  \sqrt{(1+z_2)^2-4 z_0^2}\cos\theta_0 \label{eq:H} \nonumber \\
&& + \frac{N U_0}{4} \left[(1+z_2)^2+4z_0^2\right]\nonumber \\
&&- J_1 \sqrt{(1-z_2)^2-4 z_1^2} \cos\theta_1   \nonumber \\
&& + \frac{NU_1}{4} \left[ (1-z_2)^2+4z_1^2\right]\nonumber \\
&& - N U_{01}\left[z_0+z_1-z_2(z_0-z_1)\right]\sin\theta_2\sin(\theta_0-\theta_1)\nonumber \\
&&+ \frac{N U_{01}}{2}\left(1-z_2^2 +4 z_0 z_1\right) \left[2+\cos\theta_2\cos(\theta_0-\theta_1) \right]\nonumber \\
&& -\Delta E z_2,
\end{eqnarray}
where $\Delta E\equiv E_1-E_0$. Note that the coordinate $\theta_N$, canonically conjugate to the total number of particles, does not appear in the Hamiltonian. In this semiclassical Hamiltonian one recognizes the terms
\begin{equation}
H_{\ell} \equiv
-J_\ell \sqrt{ (1\!\pm\! z_2)^2\!-\!4 z_\ell^2 }  \cos\theta_\ell+NU_\ell z_\ell^2
\end{equation}
 as those describing two non-rigid pendula of variables $(z_\ell,\theta_\ell)$.  These two pendula are non-trivially coupled to each other and to the third pendulum, of variables $(z_2,\theta_2)$.

The  equations of motion in terms of the new coordinates are then obtained from the relations $\dot{z}_i=-\partial H'/\partial\theta_i$ and $\dot{\theta}_i=\partial H'/\partial z_i$ as
\begin{widetext}
\begin{align}
\label{eq:eqsmotion-first}
\dot{z}_0 = \,& - J_0 \left[(1+z_2)^2-4 z_0^2\right]^{1/2} \sin\theta_0 \\
&\,+\frac{N U_{01}}{2} \left[ (1+4 z_0 z_1 -z_2^2) \cos \theta_2\sin(\theta_0-\theta_1)+ (z_0+z_1-z_0 z_2 +z_1 z_2) \sin\theta_2\cos(\theta_0-\theta_1)\right],\nonumber\\
\dot{\theta}_0 =\, & 2 z_0 \left\{N U_0+\frac{2 \cos\theta_0 J_0}{\left[(1+z_2)^2-4 z_0^2\right]^{1/2}}\right\}+N U_{01} \left\{ 2\,z_1 \left[2+\cos\theta_2 \cos(\theta_0-\theta_1)\right]-(1-z_2)\sin\theta_2 \sin(\theta_0-\theta_1)  \right\},\\
 \dot{z}_1 = \,& -J_1 \left[(1-z_2)^2-4 z_1^2\right]^{1/2}\sin\theta_1  \\
&\,-\frac{N U_{01}}{2} \left[ (1+4 z_0 z_1 -z_2^2) \cos{\theta_2}\sin(\theta_0-\theta_1)+ (z_0+z_1-z_0 z_2 +z_1 z_2) \sin \theta_2 \cos(\theta_0-\theta_1)\right],\nonumber\\
\dot{\theta}_1 =\, & 2 z_1 \left\{N U_1+\frac{2 \cos\theta_1 J_1}{\left[(1-z_2)^2-4 z_1^2\right]^{1/2}}\right\} +N U_{01} \left\{2 z_0 \left[2+\cos\theta_2 \cos(\theta_0-\theta_1)\right]-(1+ z_2) \sin\theta_2 \sin(\theta_0-\theta_1)\right\}, \\
\dot{z}_2 = \,&\frac 1 2 N U_{01} \left[ 2  (z_0+z_1-z_0 z_2 +z_1 z_2) \cos\theta_2\sin(\theta_0-\theta_1) +(1+4 z_0 z_1 -z_2^2) \sin\theta_2\cos(\theta_0-\theta_1) \right], \\
\dot{\theta}_2 = \, &  -\Delta E -\frac{ J_0 (1+z_2)\cos\theta_0}{\left[(1+z_2)^2-4 z_0^2\right]^{1/2}} + \frac{ J_1(1-z_2)\cos\theta_1}{\left[(1-z_2)^2-4 z_1^2\right]^{1/2}} \label{eq:eqsmotion-last} \\
&\,+\frac{N U_0}{2} (1+z_2)-\frac{N U_1}{2} (1-z_2)- N U_{01} \left\{z_2 \left[2+\cos\theta_2 \cos(\theta_0-\theta_1)\right]+(z_1-z_0)\sin\theta_2 \sin(\theta_0-\theta_1) \right\}.\nonumber
\end{align}
\end{widetext}
These equations can be greatly simplified in certain parameter regimes which we discuss in the following section.

\section{Bounds, regimes and fixed points}
\label{sec4}

The dynamical behavior of the system will crucially depend on the barrier height $V_0$ and the interaction strength $g_1$. In this section we identify different regimes in terms of these coefficients, find the fixed points and study their stability to gain information about the dynamics. Let us first note that, to have  localized modes, the barrier height has to satisfy $V_0 \ge E_1$. Moreover, a good definition of the ground and excited modes also requires
\begin{equation} 
\label{JellDeltaE}
J_\ell\ll\Delta E \, .
\end{equation}
Next we argue that $\Delta E$  should be larger than $N U_{01}$ in the few atom limit, where $U_{01}$, as can be inferred from Eqs.~(\ref{eq:U_l})-(\ref{eq:U_lm_l'm'}), is of the order of $U_0$~\cite{Dounas-Frazer:2007,GarciaMarch:2011,GarciaMarch:2012b}. For small interaction strengths the wave functions $\psi_{j\ell}(z)$ in a single well become the eigenfunctions of a harmonic trap of frequency $\omega$ and $\Delta E \simeq \hbar \omega$.  We have numerically solved the Gross-Pitaevskii equation for a single well to obtain the eigenstates and eigenenergies for different values of $N g_1$ as well as the single-particle energy given by Eq.~(\ref{eq:E_lm}) and confirmed that $\Delta E\simeq \hbar \omega$ in all cases of interest. When $N U_{01}$ is comparable to $\Delta E$ in a harmonic trap, then, because of the equal level spacing, one  has to consider at least one additional mode in each well.  We conclude that the
four-mode model is therefore justified if
\begin{equation}
  \chi_{01} \equiv \frac{N U_{01}}{\Delta E} \ll 1 \, . \label{condU01}
\end{equation}

The fixed points  $z_i^0,\theta_i^0$ $(i=0,1,2)$ of the global dynamical system can be obtained when all the conditions
$\dot{z}_i=\dot{\theta}_i=0$ are met for Eqs.~(\ref{eq:eqsmotion-first})-(\ref{eq:eqsmotion-last}) simultaneously. We find $z_0^0=z_1^0=0$,
$\theta_i^0=k_i \pi$, where $k_i$ takes values 0 and 1, and
\begin{equation}
z_2^0 = \frac{2\Delta E +2 (-1)^{k_0} J_0 -2 (-1)^{k_1} J_1 +N(U_1-U_0)}{N\{U_0+U_1-2U_{01}[2+(-1)^{k_0+k_1+k_2}]\}}.\label{eq:stationary}
\end{equation}
These eight fixed points correspond to an equal balance between the right and left populations for each mode, while a number difference exists between the
ground and excited modes.
Due to the requirement $|z_2^0|\leq1$, those fixed points do not always exist. Indeed, for $NU_{01}\lesssim \Delta E$ they occur at $|z_2^0|\gtrsim 1$, which is unphysical.

The condition~(\ref{condU01}) has important implications on the dynamics. By numerically solving (\ref{eq:eqsmotion-first})-(\ref{eq:eqsmotion-last}) we find that in all the physical regimes discussed below, for which it is always fulfilled,  $z_2$ stays at its initial value, while $\theta_2$ grows unbounded, resembling the self-trapping scenario described above for these variables. If one approximates $z_2$ as constant and solves Eq.~(\ref{eq:eqsmotion-last}) for a fixed point in the other two degrees of freedom, one finds an analytical expression for $\theta_2$ which approximately grows linearly with time, agreeing with the numerical solution for many periods of oscillation. Nevertheless, for long enough times (many typical oscillations of $z_1,z_2$), the effect of the excited modes becomes non-negligible, and the above mentioned analytical result for $\theta_2$ deviates from the
numerical one. Therefore, the results that follow will be valid for a large  but finite   number of  oscillations.

Inspection of Eq. (\ref{eq:eqsmotion-last}) suggests that the behavior of
$\dot{\theta}_2$ depends on how  $\Delta E$ compares to the interaction  and hopping energies. From the conditions (\ref{JellDeltaE}) and (\ref{condU01}), we note that $\theta_2(t)\simeq\Delta E t$, as expected from the previous paragraph.

Equations (\ref{eq:eqsmotion-first})-(\ref{eq:eqsmotion-last}) can be simplified by assuming that $z_2$ is constant and $\theta_2$ grows unbounded, which permits to average out the terms proportional to $\sin (\theta_2)$ and
$\cos (\theta_2)$. The resulting equations read
\begin{align}
   \dot{z}_\ell=& - J_\ell \sin\theta_\ell \sqrt{(1+(-1)^\ell z_2)^2-4z_\ell^2},
   \label{eq-zell}\\
   \dot{\theta}_\ell=& 2 z_\ell \left\{N U_\ell  + \frac{2J_\ell \cos\theta_\ell}{\sqrt{[1+(-1)^\ell z_2]^2-4z_\ell^2}}\right\}+4NU_{01} z_{\ell'}, \label{eq-thell}
\end{align}
where $\ell=0,1$ and $\ell ' \neq \ell$. Equations~(\ref{eq-zell})-(\ref{eq-thell}) are similar to those found when considering a mixture of ultracold bosons  in double-well potentials~\cite{Xu:2008,JuliaDiaz:2009,Satija:2009,Qiu:2010,Mazzarella:2010}. One important difference is that, while here $U_{01}$, $U_0$ and $U_1$ are comparable, in two component systems, the equivalent of $U_{01}$ (which is the inter-species interactions) can be tuned in an experiment.

We stress here that the numerical results below are obtained with the full set of equations of motion~(\ref{eq:eqsmotion-first})-(\ref{eq:eqsmotion-last}), and Eqs.~(\ref{eq-zell})-(\ref{eq-thell}) are only used as a simplified model to gain physical insight and derive analytical results valid in some regimes.

In two-mode descriptions of the double-well condensate, the parameter  $N U/2J$ characterizes the different dynamical regimes~\cite{Milburn:1997,Smerzi:1997,Sols:1999b,Legget:2001}. We can define two
analogous quantities for the four-mode model:
\begin{equation}
  \chi_\ell \equiv \frac{N U_\ell}{2 J_\ell}.
\end{equation}
For weak enough interactions, when the localized wavefunctions approach the solutions of the harmonic oscillator,  one can analytically show that $J_1> J_0$ and $U_1= (3/4) U_0$~\cite{Dounas-Frazer:2007,GarciaMarch:2011,GarciaMarch:2012b}. For larger interactions, $U_1$ remains of order $U_0$ and    the inequality $J_1> J_0$ continues to apply. In this limit the condition $\chi_1 <\chi_0$ is therefore satisfied.

In the following we identify the various dynamical regimes defined by the values of $\{\chi_\ell\}$. To discuss these regimes we use the simplified model~(\ref{eq-zell})-(\ref{eq-thell}) and  we present numerical checks of our results using the full equations of motion~(\ref{eq:eqsmotion-first})-(\ref{eq:eqsmotion-last}). 

\subsection{Rabi regime}

In the Rabi regime the tunneling strengths $J_\ell$ dominate over the interactions. It is characterized by
\begin{equation}
   \chi_1 < \chi_0 < 1.
\end{equation}
For vanishing interactions, the system dynamics is equivalent to that of two uncoupled non-rigid pendula~\cite{Javanainen:1987,Milburn:1997,Smerzi:1997}. Macroscopic  tunneling is predicted for any initial population imbalance. The fixed points of the system are given by
\begin{equation}
   z_\ell^0=0,
\end{equation}
and are stable for both $\theta_\ell=0$ and $\pi$.

\subsection{Mixed regime}

When the interaction strength grows, the system enters the mixed regime, characterized by
\begin{equation}
  \chi_1 < 1 < \chi_0\,,
\end{equation}
which is specific of the four-mode model.
Here, the ground modes may experience self-trapping depending on the initial conditions. The fixed points of the excited modes remain at $z_1^0=0$ and stable. The fixed point $z_0^0=0$ at $\theta_0^0=0$ remains, but the one at $\theta_0^0=\pi$ now splits into three, namely
\begin{equation}
z_0^0=0,
\end{equation}
which is unstable, and
\begin{equation}
z_0^\pm=\pm \sqrt{1-\left(\frac{1+ z_2}{\chi_0}\right)^2},
\end{equation}
which are stable and describe self-trapping dynamics. This is the pitchfork bifurcation discussed in Refs. ~\onlinecite{Smerzi:1997,Milburn:1997,Zibold:2010}.

\subsection{Josephson regime}
\label{subsec:jos}

The Josephson regime is characterized by
\begin{equation}
1 < \chi_1 < \chi_0,
\end{equation}
and can exhibit self-trapping in both levels. We will see that the non-zero value of $U_{01}$ introduces a dependency of the self-trapping threshold on the mode populations.

\begin{figure}
\begin{center}
\includegraphics[width=0.45\textwidth]{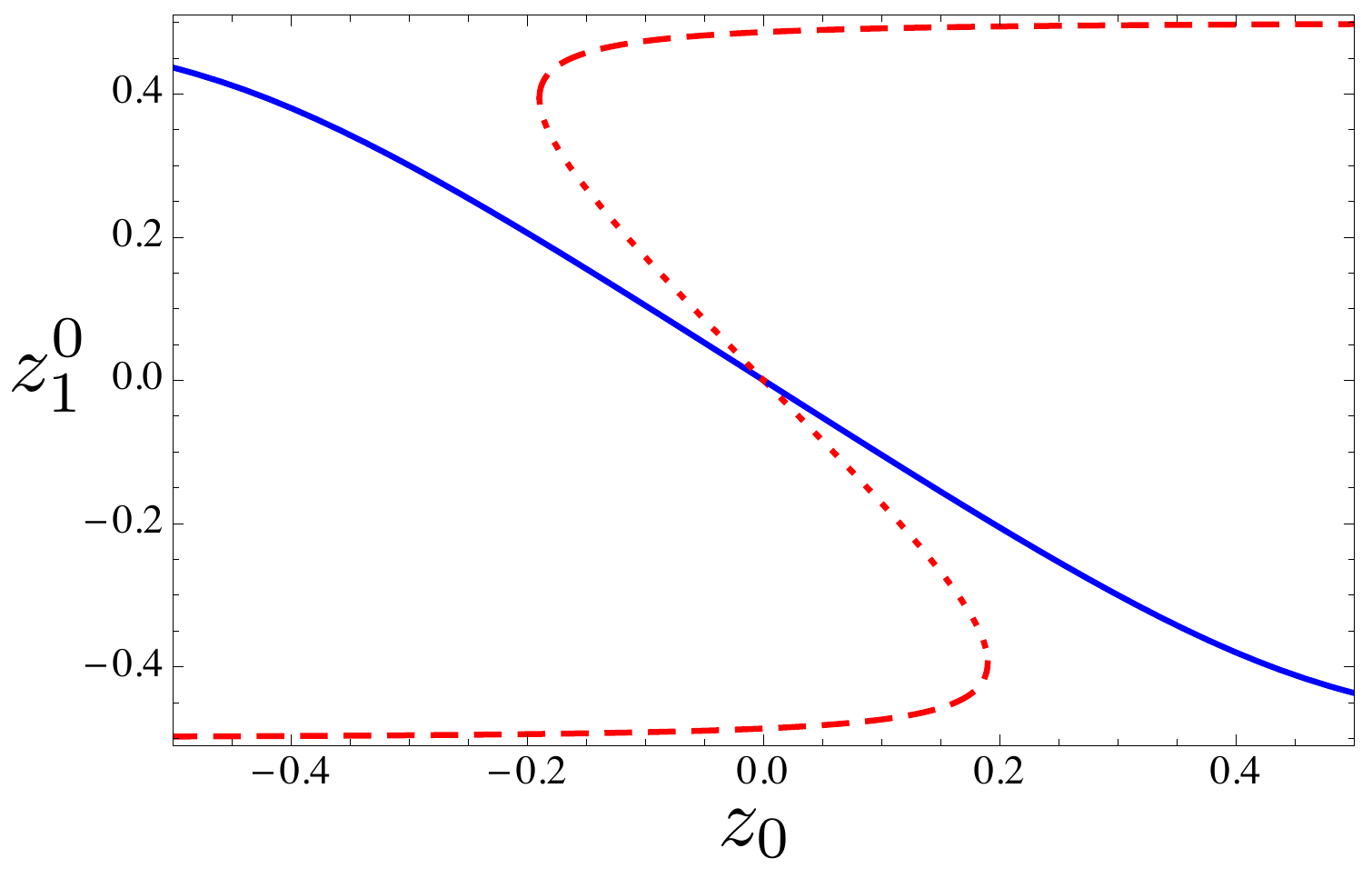}
\caption{(Color online) {\it Effective fixed points of $z_1$ in the Josephson regime for half the atoms excited} ($z_2=0$). Plotted as a function of the initial (and practically constant) value of $z_0$, which is self-trapped or tunnels slowly in this regime. The solid blue line represents the values of the effective fixed point  $z_1^0$, always with $\theta_1^0=0$. When the non-excited atoms are mostly localized initially in one well [$z_0(0)\ne0$],  the excited atoms are mostly localized in the other well, $z_1^0\ne0$. The red lines are the effective fixed points $z_1^0$ with  $\theta_1^0=\pi$. The dashed(dotted) line contains stable(unstable) effective fixed points. Above a certain value of $|z_0(0)|$, two of the effective fixed points with $\theta_1^0=\pi$ merge and disappear.  \label{fig:2.5}}
\end{center}
\end{figure}

In this regime, the fixed points are $z_0^0=z_1^0=0$,  which are stable for $\theta_0^0=\theta_1^0=0$ and unstable for $\theta_\ell^0=\pi$. We note however that, if the atoms in the ground modes are predominantly  trapped in one well, thus keeping $z_0$ nonzero and almost constant, one can find  points where $z_1$ and $\theta_1$  also remain constant  for many oscillation periods.  When the atoms in the ground modes are not self-trapped and $z_0$ oscillates, the typical oscillation frequencies of $z_0$ are much slower than those of $z_1$, since $J_1>J_0$. Thus,  we can also find  values at which   $(z_1,\theta_1)$ remain constant for times shorter than one oscillation period of the ground mode. Since in both cases ($z_0$ trapped or slowly oscillating) these points  behave as fixed points for many $z_1$ oscillations, they can be referred to as {\it effective fixed points}  with $z_1^0\ne0$. We can obtain these solutions $z_1^0$ by numerically  solving for the roots of Eq.~(\ref{eq-thell}), assuming $z_0$ 
constant.

In Fig.~\ref{fig:2.5} we show the solutions when half of the atoms are excited ($z_2 =0$). There are four effective fixed points of  $z_1$ for an initial $z_0(0)=0$: two effective fixed points at  $z_1^0=0$, one stable at $\theta_1^0=0$ and one unstable at  $\theta_1^0=\pi$, and two stable effective fixed points at $z_1^0\neq 0$ at $\theta_1^0=\pi$. They are shifted when the initial $z_0(0)$ is changed, as shown in Fig.~\ref{fig:2.5}. In particular, the two stable effective fixed points for  $\theta_1^0=\pi$ for which $z_1^0$ has the opposite sign of the initial $z_0(0)$,  approach each other as $|z_0(0)|$ grows until they reach the critical value
\begin{equation}
   |z_1^c|=\frac 1 2 \sqrt{1-[(1-z_2)/\chi_1]^{2/3}}.
\end{equation}
At larger values of $|z_0(0)|$ these two solutions become imaginary and only the other two  effective fixed points remain, one at $\theta_1^0=0$ (with opposite sign of $z_0(0)$) and one at $\theta_1^0=\pi$ (with the same sign as $z_0(0)$). This dynamical behavior is similar to that obtained for two bosonic species in a double well~\cite{Xu:2008,Mele-Messeguer:2011}. The effective fixed points shown in Fig.~\ref{fig:2.5} are confirmed  in Sec.~\ref{sec7} by numerically simulating  the full equations of motion~(\ref{eq:eqsmotion-first})-(\ref{eq:eqsmotion-last}).

This existence of effective fixed points is due to the repulsive interaction between the clouds $(U_{01}>0)$ since the term $4NU_{01} z_{0}$ in Eq.~(\ref{eq-thell}) for $\ell=1$ can be interpreted as an effective asymmetry of the double well potential, as  $z_{0}$ is  approximately constant for the time frame considered. Then,  the atoms of the ground level increase the total potential energy in one of the wells, causing the atoms in the excited modes to be trapped in the well with fewer atoms in the ground modes.

\begin{figure}
\begin{center}
\includegraphics[width=0.45\textwidth]{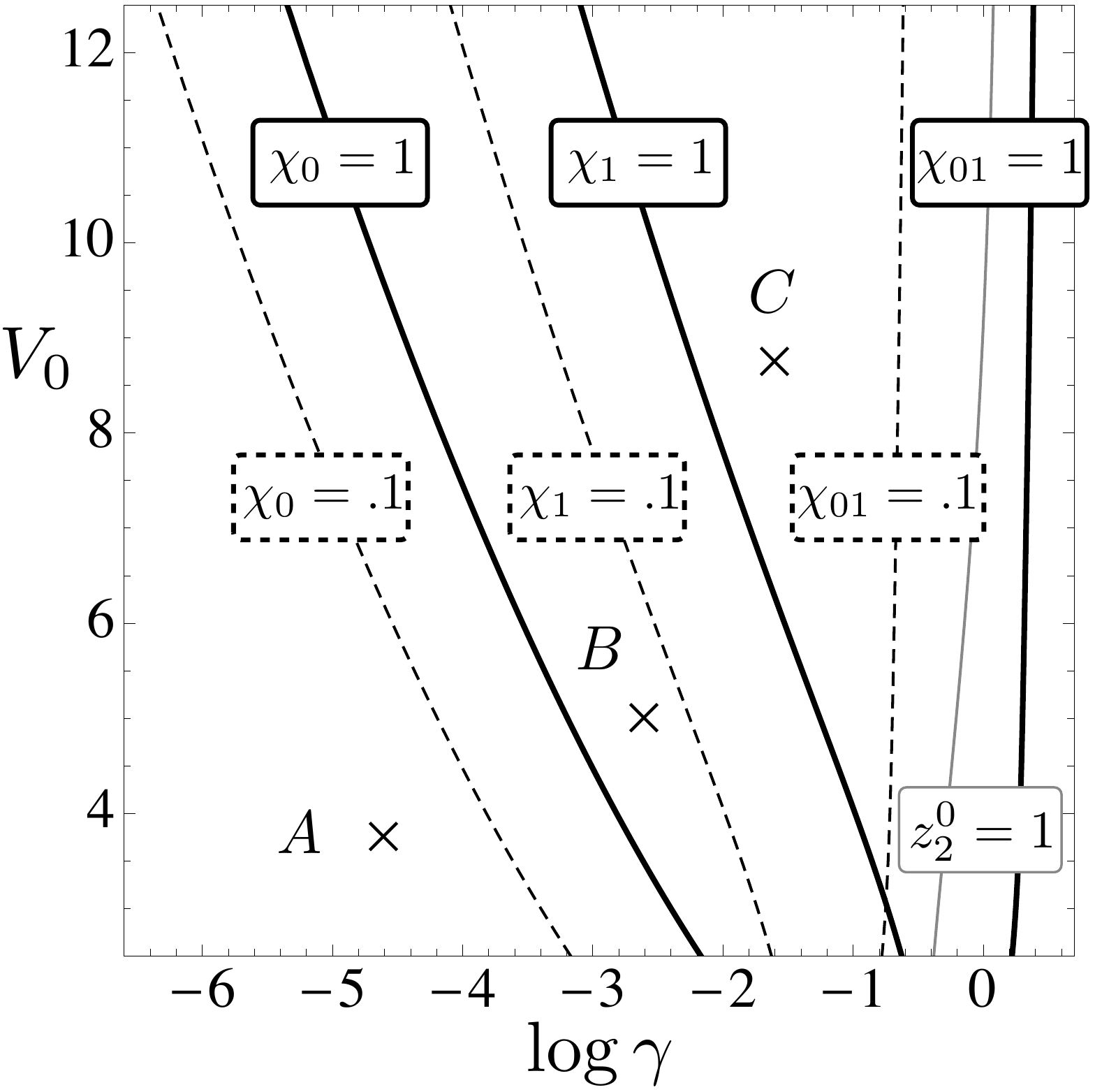}
\caption{{\it Representation of the boundaries between the different regimes in the $(V_0,\gamma)$ plane, for the Duffing potential.} See Eqs.~(\ref{eq:Duffing}), (\ref{eq:U_lm_l'm'}), and~(\ref{eq:gamma}). Thick solid lines represent the locus of points where $\chi_\ell=1$ and $\chi_{01}=1$. The Rabi regime occupies the left of the line $\chi_0=1$. The mixed regime occurs between lines $\chi_0=1$ and $\chi_1=1$. The Josephson regime occupies the right of the line $\chi_1=1$.  The shifted, dashed lines are the positions where the same quantities are 0.1. The area above the horizontal axis is the $V_0>E_1$ regime. The three cross-marked points  on the graphs ($A$, $B$, $C$) are the values chosen to investigate the dynamics in the Rabi, mixed and Josephson regimes. Finally, the right hand side of the thin solid line is the region where fixed points for the whole system exist. This line is close to the line $\chi_{01}=1$ which marks the limit of validity of the semiclassical model. } \label{fig:regimes}
\end{center}
\end{figure}

\subsection{Fock regime}
\label{subsec:fo}

Finally, the Fock regime is reached when
$$\chi_\ell\gg N^2.$$
In this limit, the relative phase between the atoms in each well is random, and the coherence between both wells is lost. The semiclassical approach is no longer valid. In the following we assume $\chi_\ell\ll N^2$, i.e.,~we keep the analysis restricted to the Josephson, Rabi, or mixed regime.

\section{Dynamics of the four-mode model}
\label{sec6}

\begin{figure}
\begin{center}
\includegraphics[width=0.45\textwidth]{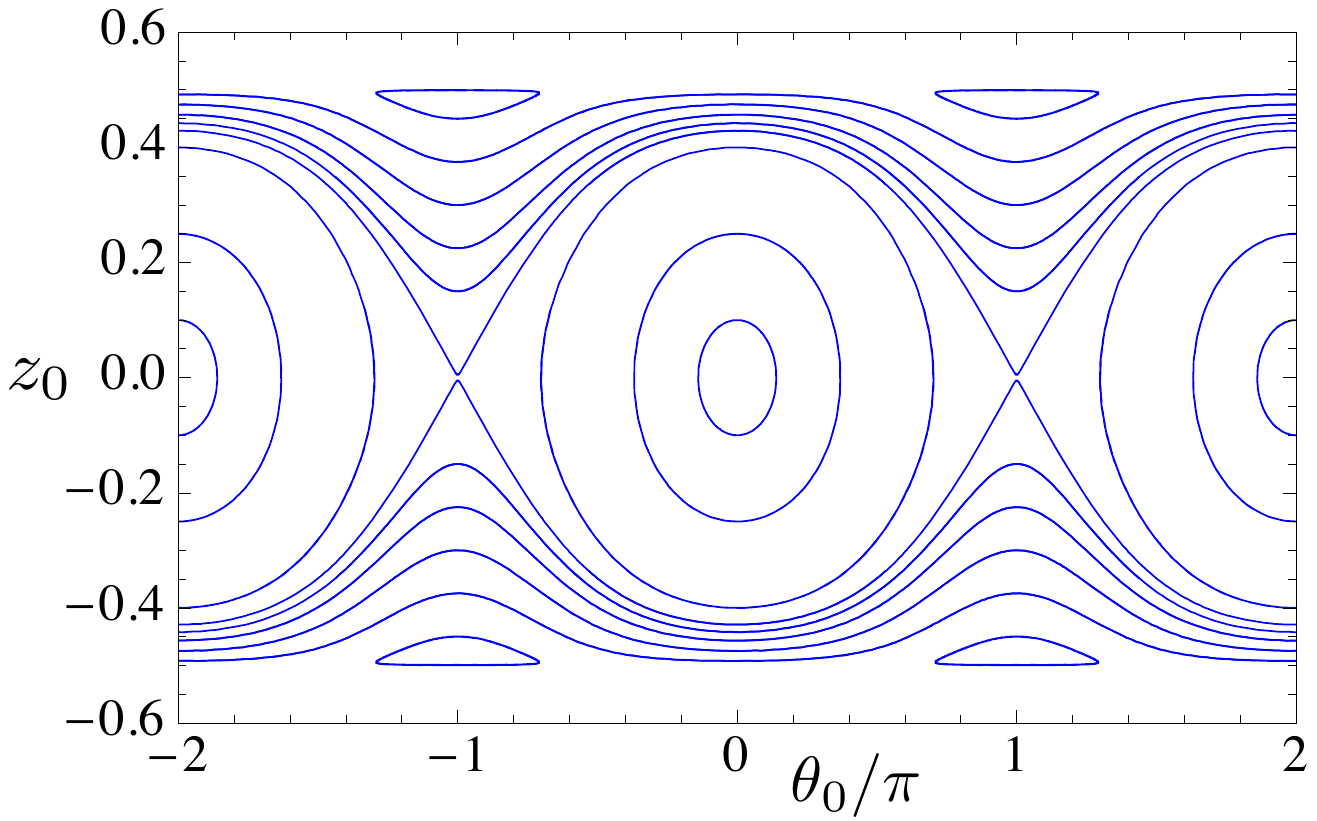}
\caption{(Color online) {\it Phase plane for $(z_0,\theta_0)$ in the mixed regime when half the atoms are excited.}  All trajectories start at $z_1(0)=\theta_1(0)=0$, which is a fixed point for the excited mode. Macroscopic quantum tunneling occurs for small initial values of $z_0$, but for larger $z_0(0)$ the system shows self-trapping in $z_0$. The fixed point of the Rabi regime at $z_0^0=0, \theta_0^0=\pi$ has bifurcated in this mixed regime into three fixed points: one unstable at the same position and two stable with $z_0^0\neq 0$ and $\theta_0^0=\pi$. For $z_1$ the fixed points are those of the Rabi regime, namely, $z_1^0=0$ and $\theta_1^0=0, \pi$, all of them stable.
\label{fig:b}}
\end{center}
\end{figure}

In the following we explore the regimes identified in Sec.~\ref{sec4} by solving the equations of motion~(\ref{eq:eqsmotion-first})-(\ref{eq:eqsmotion-last}) numerically. To reduce the number of parameters of the system to only two, we introduce
\begin{equation}
\label{eq:gamma}
\gamma \equiv N \frac{g_1}{2},
\end{equation}
as $N$ and the interaction coefficients $U_\ell$ and $U_{01}$  [see Eqs.~(\ref{eq:U_l}) and (\ref{eq:U_lm_l'm'})]
always appear as a product in Hamiltonian~(\ref{eq:H}).
In Fig.~\ref{fig:regimes} we show the $(V_0,\gamma)$ plane
separated into the different regimes by the locus of points satisfying one of the four conditions:
$\chi_\ell=1$, $\chi_{01}=1$, $\chi_\ell=0.1$, $\chi_{01}=0.1$. We stress that this figure is obtained after calculating the single particle eigenfunctions with a Duffing
potential and solving numerically the integrals (\ref{eq:J_lm})-(\ref{eq:U_lm_l'm'}) to obtain the parameters of the problem.

In Fig.~\ref{fig:regimes} we also plot the curve satisfying $z_2^0=1$ from Eq.~(\ref{eq:stationary}). To this end, for each pair $(V_0,\gamma)$, we solve the integrals (\ref{eq:J_lm})-(\ref{eq:U_lm_l'm'}) and substitute into Eq.~(\ref{eq:stationary}) made equal to 1. To the right of this curve we have interacting regimes with fixed points where the six variables of the three pendula remain constant. These  fixed points fall near the curve $ \chi_{01}= 1$, which marks the limit of validity of  our model.

We choose three sets of parameters $(V_0,\gamma)$ that show representative dynamics for each regime. These are the points $A, B, C$  of Fig.~\ref{fig:regimes}, corresponding to the Rabi, mixed, and Josephson regimes, respectively. For all the cases discussed below, we find $ \chi_{01}\ll 1$. We have numerically verified that $z_2$ oscillates for many periods with a small amplitude around its initial value with a frequency at least two orders of magnitude higher than the $z_1$-oscillations, while   the phase  $\theta_2$ grows linearly.

Point $A$ in Fig.~\ref{fig:regimes} ($V_0=3.75$, $\gamma=2.5\times10^{-5}$) corresponds the Rabi regime, as we find $\chi_0 \simeq  1.3 \times
10^{-2}$, $\chi_1 \simeq 6.6 \times 10^{-6}$ and $\chi_{01} \simeq 2.1 \times 10^{-4}$. As expected, $z_0$ and $z_1$ exhibit oscillations around zero (see Sec.~\ref{sec4}).

Point $B$ in Fig.~\ref{fig:regimes} ($V_0=5$, $\gamma=2.5\times10^{-3}$) corresponds to the mixed regime, for which we find  $\chi_0 \simeq 3.9 $, $\chi_1 \simeq 4.5\times 10^{-2}$ and $\chi_{01} \simeq 6.2 \times 10^{-4}$.
Here $z_{\ell}=\theta_{\ell}=0$ is a fixed point both for $\ell=0,1$. In Fig.~\ref{fig:b} we show the phase portrait of $(z_0,\theta_0)$ for $z_1(0)=\theta_1(0)=0$ and $z_2(0)=0$. Note that, according to Eq.~(\ref{eq:boundsforzell}), $|z_0|<1/2$. Throughout the entire  numerical evolution,  $z_1$ and $\theta_1$  do not depart from their initial value of 0 perceptibly ($z_1$ shows oscillations of the order of $10^{-2}$). This allows us to understand the phase portrait of $z_0,\theta_0$ as if $z_1,\theta_1$ were constant. Some trajectories display effective self-trapping behavior. Finally, we point out that, for initial conditions with $z_1(0)\ne0$, the atoms in the excited modes always show  oscillatory behavior similar to that of the Rabi regime in the two-mode setting (not shown).

To understand the dynamics around the stationary point $z_\ell^0=\theta_\ell^0=0$ we can linearize Eqs.~(\ref{eq-zell})-(\ref{eq-thell}) as
\begin{align}
  \dot{z}_\ell=& -\theta_\ell J_\ell [1+(-1)^\ell z_2],
  \label{eq-linz}\\
  \dot{\theta}_\ell=& 2 z_\ell  U_\ell+4  z_\ell J_\ell/[1+(-1)^\ell z_2] +4U_{01}z_{\ell '}. \nonumber
\end{align}
We stress again that these equations are valid only for a finite number of oscillations, but are good guides to interpret the numerical simulations. Equations~(\ref{eq-linz}) can be interpreted as two coupled oscillators with normal-mode frequencies
\begin{equation}
  \omega_\pm^2=\frac{1}{2}\left[\omega_0^2+\omega_1^2\pm\sqrt{(\omega_0^2-\omega_1^2)^2 +64(1- z_2^2)J_0 J_1 U_{01}}\right],
\label{Eq:phaseshift}
\end{equation}
where the  frequencies of the two linearized two-mode models are
\begin{equation}
\omega_{\ell}^2/J_\ell= 2NU_\ell[1+(-1)^\ell z_2]+4J_\ell. 
\end{equation}
 We note the interesting relationship
\begin{equation}
\omega_-<\omega_0<\omega_1<\omega_+ \, .
\end{equation}

\begin{figure}
\includegraphics[width=0.48\textwidth]{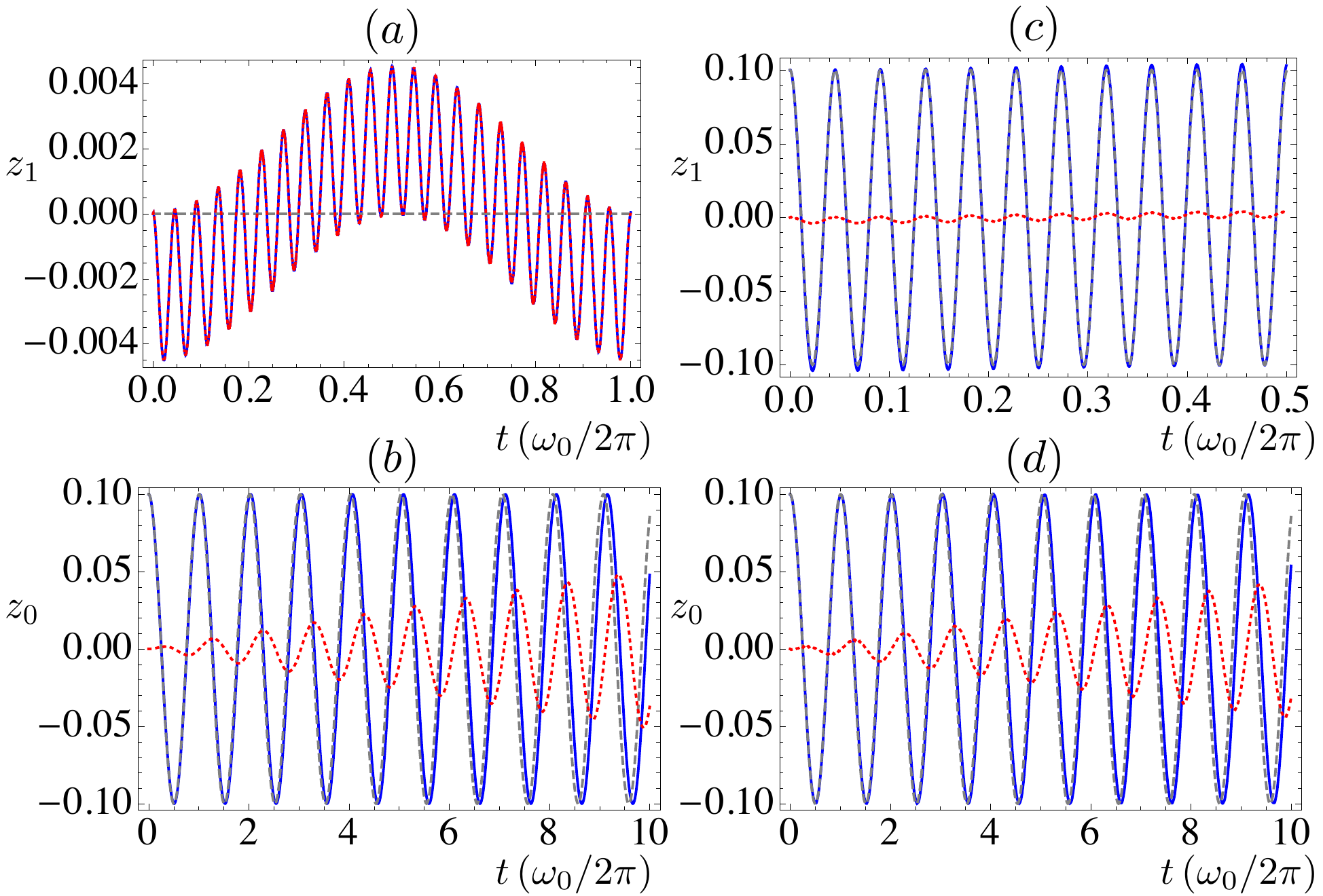}
\caption{(Color online)  {\it Comparison between the two- and four-mode models around the effective fixed points in the mixed regime with $20\%$ of atoms excited.} The blue solid (gray dashed) lines represent the  calculations with the four (two) mode model, while the dotted red solid lines are the difference between both models. In $(a)$ and $(b)$, $z_0$ oscillates around a fixed point, while  $z_1$ starts at a fixed point and is pushed out of equilibrium by the atoms in the ground modes. In $(c)$ and $(d)$ both $z_0$ and $z_1$ are oscillating around a fixed point. In both cases, the presence of the excited modes induces a decrease of the frequency of the ground modes, which is a first approximation to the effect of dissipation in realistic models with more modes. \label{fig:5}}
\end{figure}

\begin{figure}
\includegraphics[width=0.48\textwidth]{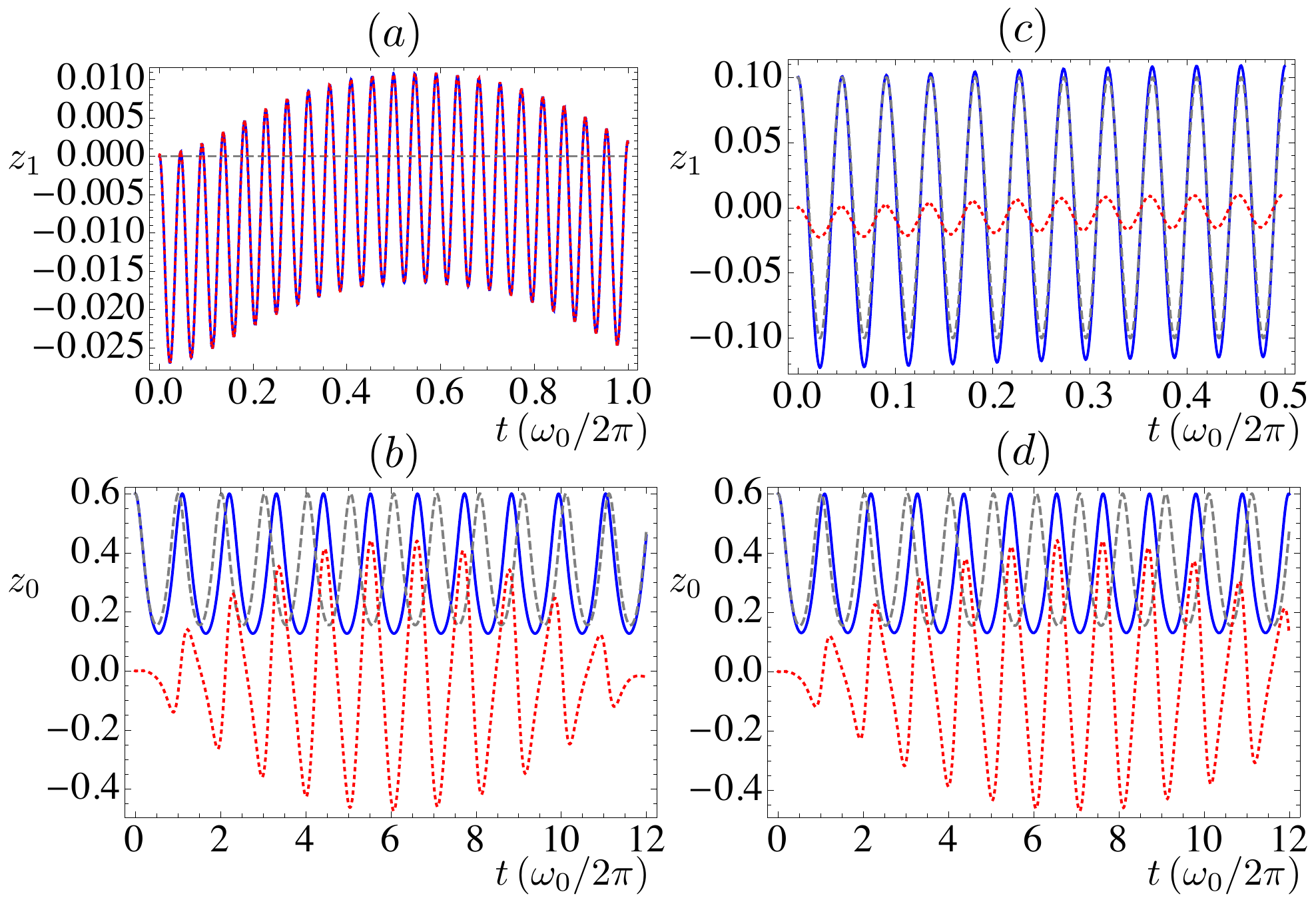}
\caption{(Color online) {\it Comparison between the two- and four-mode models for $z_0$ self-trapped in the mixed regime, with $20\%$ of atoms excited.} Same conventions as in Fig.~\ref{fig:5} but with  $z_0$ undergoing self-trapped dynamics. We again see a dragging of the excited modes by the ground atoms as well as frequency shifts for the ground atoms, which are stronger than those in Fig.~\ref{fig:5}.
\label{fig:6}}
\end{figure}

In Figs.~\ref{fig:5}a-b we present the time evolution of $z_1$ and $z_0$, respectively, for the initial conditions $z_1(0)=\theta_1(0)=0$ and $z_0(0)=0.1$, $\theta_0(0)=0$. Here, we assume that $20\%$ of the atoms are in the excited modes ($z_2=0.6$) and all other parameters correspond to point $B$ in Fig.~\ref{fig:regimes}. This setup may mimick a condensed cloud close to equilibrium  with a small depletion. We plot
the results of the two-mode approach [$U_{01}=0$ in Eqs.~(\ref{eq-zell})-(\ref{eq-thell})], the four-mode case [obtained from  the full equations of motion~(\ref{eq:eqsmotion-first})-(\ref{eq:eqsmotion-last})], and their difference. As expected, for the Rabi regime (not shown), we find no perceptible difference between the two-mode and the four-mode models for a large number of oscillating periods. Conversely, in the mixed regime, the presence of the excited modes induces a phase shift, similar to the one obtained analytically in Eq.~(\ref{Eq:phaseshift}) for small perturbations around fixed points. In particular, the intra-well inter-level interactions $U_{01}$ cause the oscillation frequency of the ground modes to decrease, when compared with the two mode case. Note that the atoms in the ground mode drag the excited atoms slightly out of the equilibrium, leading to shifted oscillations in the  excited population (see Fig.~\ref{fig:5}a). In Figs.~\ref{fig:5}c-d we show a 
similar case, where
the atoms in the excited modes do not start in a fixed point, that is $z_1(0)=0.1$. The presence of the ground mode slightly modifies the amplitude of the 
$z_1$ oscillations. On the other hand, the population of the excited mode induces a frequency shift in the oscillations of $z_0$.

When the initial conditions for the ground modes correspond to self-trapping dynamics, the shift obtained in the oscillation frequencies is larger, and the dragging of the excited modes by the ground modes stronger. This is shown in Figs.~\ref{fig:6}a-b, where initially $z_1(0)=\theta_1(0)=0$ corresponds to a fixed point, and $z_0(0)=0.6$, $\theta_0(0)=0$. If the atoms in the excited modes are not at a fixed point but close to it [e.g. $z_1(0)=0.1$], the shift in frequencies also  occurs for $z_0$, as shown in Fig.~\ref{fig:6}c-d. The  shift and dragging effects are due to the repulsion between the atoms in the different modes.  Notice that the atoms in the ground modes oscillate more slowly than the excited ones.   Extrapolating this slowing down to a model with a large number of excited modes oscillating with non-commensurate frequencies, this effect will not be periodic.  Therefore, the ground state atoms experience a force which induces a negative frequency shift, which can be 
interpreted as 
the onset of
dissipation.

\begin{figure}
\begin{tabular}{c}
\includegraphics[width=0.4\textwidth]{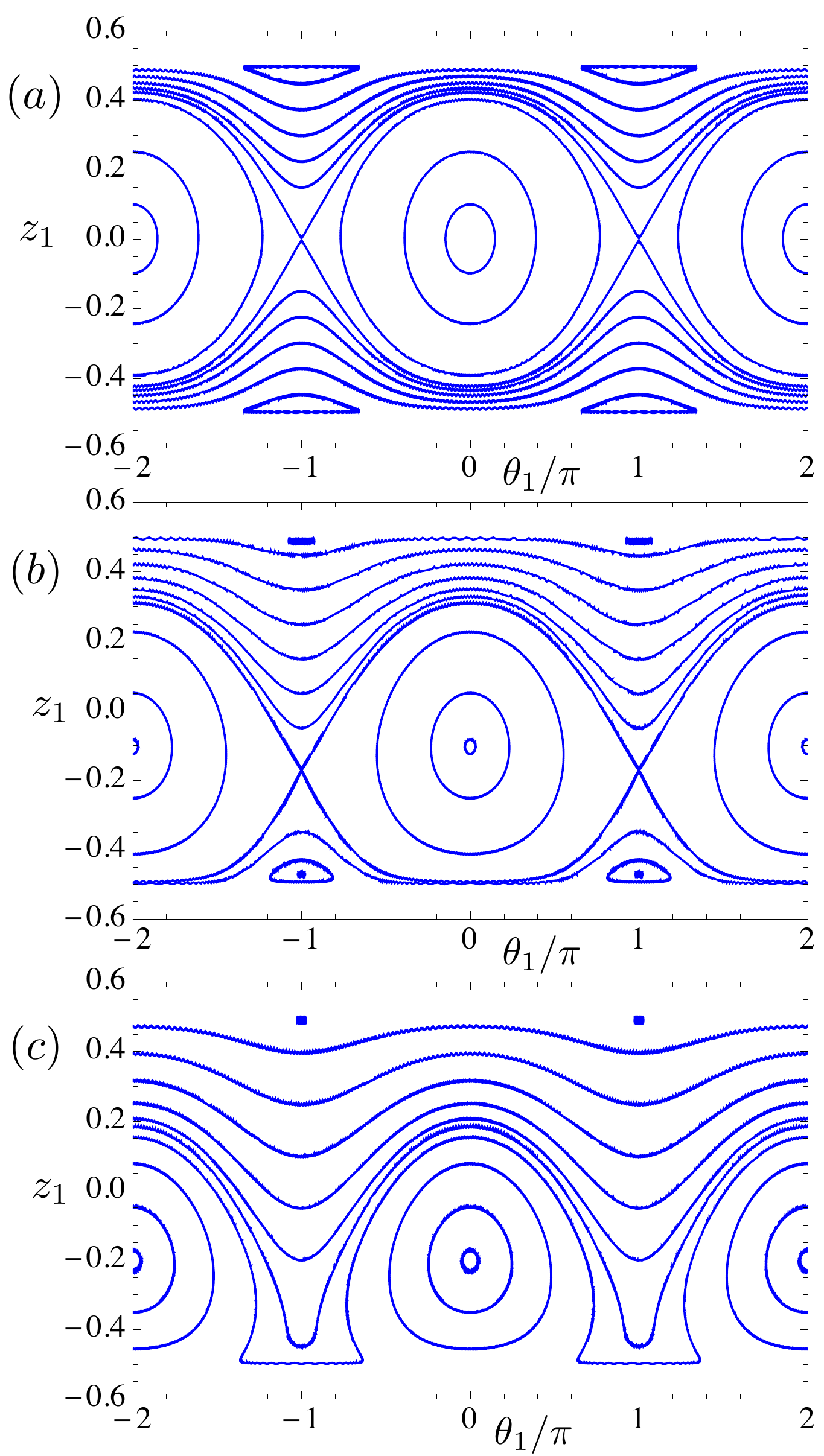}
\end{tabular}
\caption{ (Color online)  {\it Dynamics in the Josephson regime with half the atoms excited.} Phase plane portraits of $z_1,\theta_1$ with $\theta_{0}(0)=0$. In that regime and for time intervals of a few oscillations of $z_1$, the value of $z_0$ does not vary appreciably. In $(a)$, where $z_{0}(0)=0$,  we find a phase portrait very similar to that shown in Fig.~\ref{fig:b}. In $(b)$, where $z_{0}(0)=0.1$, the effective fixed points can be seen to be shifted because of the population imbalance in the ground state. In $(c)$, where  $z_{0}(0)=0.2$,  some of the effective fixed points have completely disappeared.}  \label{fig:9}
\end{figure}

\section{Manipulation of the higher modes through the population of the ground modes}
\label{sec7}

When the interactions are increased further the system enters the Josephson regime and the strong coupling between the two levels changes the fixed points, as discussed in Sec.~\ref{subsec:jos}. For the point $C$ in Fig.~\ref{fig:regimes} ($V_0=8.75$ and $\gamma=2.5\times 10^{-2}$), for which $\chi_0 \simeq 600 $, $\chi_1 \simeq
4.3$ and $\chi_{01} \simeq 5.6 \times 10^{-3}$, we find that even when the initial condition for the excited modes correspond to the point $z_{1}(0)=\theta_{1}(0)=0$, the strong coupling prevents the system from being stationary.

However, the fact that  $z_0$ varies very slowly compared to the typical oscillation frequencies of $z_1$  allows us to treat $z_0$ as an effective constant for short enough time intervals. This allows us to interpret phase portraits for $(z_1,\theta_1)$ as if $(z_0,\theta_0)$ were constant.  We show in Fig.~\ref{fig:9} three such phase portraits for $(z_{1},\theta_{1})$ corresponding to $\theta_0(0)=0$,  $z_0(0)=0,0.1, 0.2$, while $z_2=0$.    From Fig.~\ref{fig:9}a it can be seen that self-trapping is now also possible for atoms in the excited level,  the dynamics of $(z_1,\theta_1)$ in this case being very similar to that of $(z_0,\theta_0)$ shown in Fig.~\ref{fig:b} (notice the similar ratios
$\chi_\ell$ in both cases).  This shows that the same effect of self-trapping previously observed in two-mode models can be achieved in the excited modes of a four-mode model.  Note also that, in this case, there are four fixed points for $z_1$, which are those shown in Fig.~\ref{fig:2.5} for the vertical line at $z_0=0$.

For small but nonzero values of $z_{0}(0)$ the atoms in the ground level  still tunnel between the wells, but we numerically observe that their oscillation frequency $\omega_0$  is at least one order of magnitude smaller than its excited counterpart, $\omega_1$.  In Fig.~\ref{fig:9}b we  show the same phase portrait for $z_0(0)=0.1$ and, as predicted in Sec.~\ref{subsec:jos}, we observe that the position of the effective fixed points for $(z_1,\theta_1)$
is gradually shifted as $z_0(0)$ is increased. The stable effective fixed point at $\theta_1^0=0$  is moved to negative values of $z_1^0$. The unstable effective fixed point at $\theta_1^0=\pi$ is also shifted downwards. The two stable effective fixed points at $\theta_1^0=\pi$ are less affected (see Fig.~\ref{fig:2.5} for reference).  In Fig.~\ref{fig:9}c one can see that, due to the larger value of $z_0(0)$, the stable point at $\theta_1^0=0$ is shifted even further. On the other hand, the two fixed points at $\theta_1^0=\pi$ with $z_1<0$ have completely disappeared (see discussion in Sec.~\ref{subsec:jos}).

This dynamical behavior is another consequence of the repulsion between the atoms in the ground and excited modes. An initial population imbalance in $z_0$ can be viewed as an effective asymmetry of the double well for $z_1$ and the equilibrium points are shifted accordingly. The fixed point associated with macroscopic tunneling for $z_1=0$ is shifted towards the less populated well and, since the time average of $z_1$ is no longer zero, the effect appears as self-trapping.
This behavior can be magnified in the  experimentally realizable case of  $z_0(0)=-z_1(0)=0.5$, i.e.,~when half of the atoms in one well are intentionally excited, while those in the other well are not. In general, by tuning the effective 1D scattering properties of the atoms and the initial populations of the atoms in an energy level, one can influence the dynamics of the atoms in the other energy level, even if on average no (or only a few) atoms are exchanged between the different energy modes.

Finally, we would like to note that, in the region of $(V_0,\gamma)$ to the right of point C in Fig.~\ref{fig:regimes} (but still satisfying $\chi_{01} < 1$), the dynamics near unstable fixed points can be chaotic for some initial conditions. In such regimes, all dynamic variables change rapidly and the phase portraits must be replaced by Poincar\'e sections. For instance, near $z_\ell=0$, $\theta_\ell=\pi$, a slight shift in the initial conditions may induce an abrupt change from quasi-periodic to chaotic behavior. Some features of this complex dynamics have been explored in the two-component analog of this problem~\cite{Qiu:2010}, which obeys equations similar to (\ref{eq-zell})-(\ref{eq-thell}). A detailed study of chaos in this system falls beyond the scope of this work and is currently being  undertaken.

\section{Conclusions}
\label{sec8}

In this work, we have shown how the addition of two (macroscopically populated) excited modes modifies the well-known dynamical scenarios of the two-mode model for a double-well condensate. We have focused on  regimes close to those found in the two-mode model. This means that we have assumed the level spacing to be greater than the energy ($U_{01}$) which, stemming from the local interaction, characterizes both the intra-well inter-level coupling and the repulsive interaction between atoms in the ground and excited modes within the same well. We have found two main effects. The first one  is that the semiclassical inclusion of the excited modes provides a simple model for the effect of  dissipation on the collective dynamics of the ground modes. An interesting result is the slowing down
of the low-lying oscillations in the
presence of excited atoms. The second result is the discovery of a rich dynamics resulting from the manipulation of the population of the ground and excited modes. This dynamics can be explored in properly designed setups, where part of the population is intentionally excited. We wish to point out that the dynamics of the four-mode model can be even richer and include chaotic behavior near some unstable fixed points.

\acknowledgments
TB and MAGM acknowledge the support by Science Foundation Ireland under Project No.~10/IN.1/I2979. MAGM acknowledges the support of a MEC/Fulbright grant and  of grants FIS2011-24154 (MINECO) and 2009SGR-1289 (Generalitat Catalunya). FS thanks the support of grants FIS2010-21372 (MINECO) and MICROSERES-CM-S2009/TIC-1476 (Comunidad de Madrid). MAGM acknowledges useful conversations with B.~Juli\'a-D\'iaz and I. Zapata. We also thank L.~D.~Carr for extensive and valuable discussions.


\begin{thebibliography}{99}

\bibitem{Browaeys:2005}
A.~Browaeys, H.~Haffner, C.~McKenzie, S.~L. Rolston, K.~Helmerson, and W.~D. Phillips, Phys. Rev. A \textbf{72}, 053605 (2005).

\bibitem{Spielman:2006}
I.~B. Spielman, P.~R. Johnson, J.~H. Huckans, C.~D. Fertig, S.~L. Rolston, W.~D. Phillips, and J.~V. Porto, Phys. Rev. A \textbf{73}, 020702 (2006).

\bibitem{Muller:2007}
T.~Muller, S.~Folling, A.~Widera, and I.~Bloch, Phys. Rev. Lett. \textbf{99}, 200405 (2007).

\bibitem{Clement:2009}
D.~Clement, N.~Fabbri, L.~Fallani, C.~Fort, and M.~Inguscio, New J. Phys. \textbf{11}, 103030 (2009).

\bibitem{Wirth:2011}
G.~Wirth, M.~Olschlager, and A.~Hemmerich, Nat. Phys. \textbf{7}, 147 (2011).

\bibitem{Olschlager:2011}
M.~Olschlager, G.~Wirth, and A.~Hemmerich, Phys. Rev. Lett. \textbf{106}, 015302
  (2011).

\bibitem{Lewenstein:2011}
M.~Lewenstein and W.~V. Liu, Nat. Phys. \textbf{7}, 101 (2011).

\bibitem{Chatterjee:2010}
B.~Chatterjee, I.~Brouzos, S.~Zollner, and P.~Schmelcher, Phys. Rev. A
  \textbf{82}, 043619 (2010).

\bibitem{JuliaDiaz:2010b}
B.~Juli\'a-D\'{\i}az, J.~Martorell, M.~Mele-Messeguer, and A.~Polls, Phys. Rev. A
  \textbf{82}, 063626 (2010).

\bibitem{Strauch:2008}
F.~W. Strauch, M.~Edwards, E. Tiesinga, C. Williams, and C.~W. Clark, Phys. Rev. A
  \textbf{77}, 050304 (2008).

\bibitem{GarciaMarch:2011}
M.~A. Garcia-March, D.~R. Dounas-Frazer, and L.~D. Carr, Phys. Rev. A
  \textbf{83}, 043612 (2011).

\bibitem{GarciaMarch:2012}
M.~A. Garcia-March and L.~D. Carr, arXiv:1203.3206 (2012).

\bibitem{Heimsoth:2012}
M.~Heimsoth, C.~E. Creffield, L.~D. Carr, and F.~Sols, New J. Phys.
  \textbf{14}, 075023 (2012).

\bibitem{Javanainen:1987}
J.~Javanainen, Phys. Rev. Lett. \textbf{57}, 3164 (1986).

\bibitem{Smerzi:1997}
A.~Smerzi, S.~Fantoni, S.~Giovanazzi, and S.~R. Shenoy, Phys. Rev. Lett.
  \textbf{79}, 4950 (1997).

\bibitem{Milburn:1997}
G.~J. Milburn, J.~Corney, E.~M. Wright, and D.~F. Walls, Phys. Rev. A
  \textbf{55}, 4318 (1997).

\bibitem{Zapata:1998}
I.~Zapata, F.~Sols, and A.~J. Leggett, Phys. Rev. A \textbf{57}, R28 (1998).

\bibitem{Raghavan:1999}
S.~Raghavan, A.~Smerzi, S.~Fantoni, and S.~R. Shenoy, Phys. Rev. A \textbf{59},
  620 (1999).

\bibitem{Ostrovskaya:2000}
E.~A. Ostrovskaya, Y.~S. Kivshar, M.~Lisak, B.~Hall, F.~Cattani, and D. Anderson, Phys. Rev.
  A \textbf{61}, 031601 (2000).

\bibitem{Mahmud:2005}
K.~W.~Mahmud, H.~Perry, and W.~P.~Reinhardt, Phys. Rev. A \textbf{71}, 023615 (2005).

\bibitem{Ananikian:2006}
D.~Ananikian and T.~Bergeman, Phys. Rev. A \textbf{73}, 013604 (2006).

\bibitem{Fu:2006}
L.B.~Fu and J.~Liu, Phys. Rev. A \textbf{74}, 063614 (2006).

\bibitem{JuliaDiaz:2010}
B.~Juli\'a-D\'{\i}az, D.~Dagnino, M.~Lewenstein, J.~Martorell, and A.~Polls, Phys. Rev. A \textbf{81}, 023615 (2010).

\bibitem{Albiez:2005}
M.~Albiez, R.~Gati, J.~F{\"o}lling, S.~Hunsmann, M.~Cristiani, and
  M.~K.~Oberthaler, Phys. Rev. Lett. \textbf{95}, 010402 (2005).

\bibitem{Levy:2007}
S.~Levy, E.~Lahoud, I.~Shomroni, and J.~Steinhauer, Nature \textbf{449}, 579 (2007).

\bibitem{Zibold:2010}
T.~Zibold, E.~Nicklas, C.~Gross, and M.~K. Oberthaler, Phys. Rev. Lett.  \textbf{105}, 204101 (2010).

\bibitem{Dalton:2012}	
B.~J.~Dalton and S.~ Ghanbari, J. Mod. Optics    \textbf{59}  287 (2012).

  \bibitem{Gertjerenken:2013b}
  B.~Gertjerenken and C.~Weiss, Phys. Rev. A \textbf{88}, 033608 (2013).

\bibitem{Jezek:2013}
D.~M.~Jezek, P.~Capuzzi, and H.~M.~Cataldo,  Phys. Rev. A \textbf{87}, 053625 (2013).

\bibitem{Mazzarella:2013}
G.~Mazzarella and L.~Dell'Anna, Eur. Phys. J. Special Topics  \textbf{217} 197 (2013).

\bibitem{Steel:1998}
M.~J. Steel and M.~J. Collett, Phys. Rev. A \textbf{57}, 2920 (1998).

\bibitem{Cirac:1998}
J.~I. Cirac, M.~Lewenstein, K.~Molmer, and P.~Zoller, Phys. Rev. A \textbf{57},
  1208 (1998).

\bibitem{Gordon:1999}
D.~Gordon and C.~M. Savage, Phys. Rev. A \textbf{59}, 4623 (1999).

\bibitem{Higbie:2004}
J.~Higbie and D.~M. Stamper-Kurn, Phys. Rev. A \textbf{69}, 053605 (2004).

\bibitem{Huang:2006}
Y.~P. Huang and M.~G. Moore, Phys. Rev. A \textbf{73}, 023606 (2006).

\bibitem{Piazza:2008}
F.~Piazza, L.~Pezz\'{e}, and A.~Smerzi, Phys. Rev. A \textbf{78}, 051601 (2008).

\bibitem{Mazets:2008}
I.~E. Mazets, G.~Kurizki, M.~K. Oberthaler, and J.~Schmiedmayer, Europhys. Lett.  \textbf{83}, 60004 (2008).

\bibitem{Carr:2010}
L.~D. Carr, D.~R. Dounas-Frazer, and M.~A. Garcia-March, Europhys. Lett. \textbf{90}, 10005 (2010).

\bibitem{Watanabe:2010}
G.~Watanabe, Phys. Rev. A \textbf{81}, 021604(R) (2010).

 \bibitem{He:2012}
 Q.~Y.~He, P.~D.~Drummond, M.~K.~Olsen, and M.~D.~Reid, Phys. Rev. A \textbf{86}, 023626 (2012).

\bibitem{Csire:2012}
G.~Csire and B.~Apagyi, Phys. Rev. A \textbf{85}, 033613 (2012).

\bibitem{Gertjerenken:2013}
B.~Gertjerenken, T.~P.~Billam, C.~L.~Blackley, C.~R.~Le Sueur, L.~Khaykovich, S.~L.~Cornish, and C.~Weiss, Phys. Rev. Lett. \textbf{111}, 100406 (2013).


\bibitem{Sols:1999b}
F.~Sols, {\em Josephson effect between Bose condensates}, in {\em Bose-Einstein Condensation
in Atomic Gases}, Proceedings of the International School of Physics Enrico Fermi (1999), M. Inguscio, S. Stringari, and C. E. Wieman, eds., IOS Press (Amsterdam, 1999).

\bibitem{Legget:2001}
A.~J. Leggett, Reviews of Modern Physics \textbf{73}, 307--356 (2001).

\bibitem{Lipkin:1965}
H.~J. Lipkin, N.~Meshkov, and A.~J. Glick, Nucl. Phys. \textbf{62}, 188 (1965).

\bibitem{Vidal:2004}
J.~Vidal, G.~Palacios, and C.~Aslangul, Phys. Rev. A \textbf{70}, 062304 (2004).

\bibitem{Masiello:2005}
D.~Masiello, S.~B. McKagan, and W.~P. Reinhardt, Phys. Rev. A \textbf{72}, 063624 (2005).

\bibitem{Streltsov:2006}
A.~I. Streltsov, O.~E. Alon, and L.~S. Cederbaum, Phys. Rev. A \textbf{73},  063626 (2006).

\bibitem{Alon:2008}
O.~E. Alon, A.~I. Streltsov, and L.~S. Cederbaum, Phys. Rev. A \textbf{77},
  033613 (2008).

\bibitem{Sakmann:2009}
K.~Sakmann, A.~I. Streltsov, O.~E. Alon, and L.~S. Cederbaum, Phys. Rev. Lett.
  \textbf{103}, 220601 (2009).


\bibitem{Bucker:2011}
R.~B{\"u}cker, J.~Grond, S.~Manz, T.~Berrada, T.~Betz, C.~Koller, U.~Hohenester, T.~Schumm, A.~Perrin, and J.~Schmiedmayer, Nat. Phys. \textbf{7}, 608--611 (2011).

\bibitem{Bucker:2012}
R.~B{\"u}cker, T.~Berrada, S.~van Frank, J.-F. Schaff, T.~Schumm, J.~Schmiedmayer, G.~J{\"a}ger, J.~Grond, and U.~Hohenester, J. Phys. B \textbf{46}, 104012 (2013).

\bibitem{Olshanii:1998}
M.~Olshanii, Phys. Rev. Lett. \textbf{81}, 938 (1998).

\bibitem{GarciaMarch:2012b}
M.~A. Garcia-March, D.~R. Dounas-Frazer, and L.~D. Carr, Front. Phys. \textbf{7}, 131--145 (2012).

\bibitem{Spekkens:1998}
R.~W. Spekkens and J.~E. Sipe, Phys. Rev. A \textbf{59}, 3868 (1999).

\bibitem{Dounas-Frazer:2007}
D.~R. Dounas-Frazer, A.~M. Hermundstad, and L.~D. Carr, Phys. Rev. Lett. \textbf{99}, 200402 (2007).

\bibitem{davismj2008}
M.~J. Davis, R.~J. Ballagh, and C.~W. Gardiner, Adv. Phys. \textbf{57}, 363 (2008).

\bibitem{Xu:2008}
X.-Q. Xu, L.-H. Lu, and Y.-Q. Li, Phys. Rev. A \textbf{78}, 043609 (2008).

\bibitem{JuliaDiaz:2009}
B.~Juli\'a-D\'{\i}az, M.~Guilleumas, M.~Lewenstein, A.~Polls, and A.~Sanpera, Phys. Rev. A \textbf{80}, 023616 (2009).

\bibitem{Satija:2009}
I.~I. Satija, R.~Balakrishnan, P.~Naudus, J.~Heward, M.~Edwards, and C.~W. Clark, Phys. Rev. A \textbf{79}, 033616 (2009).

\bibitem{Qiu:2010}
H.~Qiu, J.~Tian, and L.-B. Fu, Phys. Rev. A \textbf{81}, 043613 (2010).

\bibitem{Mazzarella:2010}
G.~Mazzarella, M.~Moratti, L.~Salasnich, and F.~Toigo, J. Phys. B \textbf{43}, 065303 (2010).

\bibitem{Mele-Messeguer:2011}
M.~Mele-Messeguer, B.~Juli\'a-D\'{\i}az, M.~Guilleumas, A.~Polls, and A.~Sanpera, New J. Phys. \textbf{13}, 033012 (2011).

\end{thebibliography}

\end{document}